\documentclass[twocolumn,showpacs,superscriptaddress,amsfonts,amssymb,pre,aps]{revtex4-1}
\usepackage{graphicx}
\begin{document}
\title{Double resonance in the infinite-range quantum Ising model}
\author{Sung-Guk Han}
\affiliation{Department of Physics and BK21 Physics Research Division,
Sungkyunkwan University, Suwon 440-746,
Korea}
\author{Jaegon Um}
\affiliation{School of Physics, Korea Institute for Advanced Study, Seoul 130-722,
Korea}
\author{Beom Jun Kim}
\email[Corresponding author: ]{beomjun@skku.edu}
\affiliation{Department of Physics and BK21 Physics Research Division,
Sungkyunkwan University, Suwon 440-746,
Korea}
\begin{abstract}
We study quantum resonance behavior of the infinite-range kinetic Ising model
at zero temperature. Numerical integration of the time-dependent Schr\"odinger
equation in the presence of an external magnetic field in the $z$ direction
is performed at various transverse field strengths $g$.  It is
revealed that two resonance peaks occur when the energy gap matches the external
driving frequency at two distinct values of $g$, one below and the other above
the quantum phase transition.  
From the similar observations already made in classical systems with phase transitions, 
we propose that the double resonance peaks should be a generic feature
of continuous transitions, for both quantum and classical many-body systems. 

\end{abstract}

\maketitle

\section{Introduction}

A noise is often considered as a nuisance for a system to display any ordered behavior, 
and thus the weaker the better for the performance of the system. However, for the last decades, 
a lot of researchers have revealed that this is not always the case and that there exist
a class of systems in which the intermediate strength of noise can help the system to show
the best coherence with an external periodic driving. This surprising phenomenon was termed as
the stochastic resonance (SR) due to its stochastic nature~\cite{Gammaitoni}.
The phenomenon of SR has been found in the fields of physics
and earth science, as well as in biology: 
the periodically recurrent ice ages, 
the bistable ring laser, superconducting quantum interference device, 
human vision and th3 auditory system, and
the feeding mechanism of paddle fish, to list a few~\cite{Gammaitoni,ExofSR}. 
The occurrence of the SR is properly explained by the time-scale matching condition: 
the coherence between the system's response and the external driving
becomes strongest when the stochastic time scale inherent in the system 
matches the time scale provided by the external driving. In a simple classical system 
of a few degrees of freedom making contact with a thermal reservoir, the intrinsic time scale 
is given by the monotonically decreasing function of the exponential thermal activation form.
Accordingly, the above mentioned time-scale matching condition can only be satisfied at a
single temperature~\cite{Gammaitoni}. 
The time-scale matching condition was later extended to the classical
statistical mechanical systems with continuous phase transitions such as
the globally coupled, i.e., infinite-range kinetic Ising model~\cite{Kim}.
It has been shown that the nonmonotonic behavior of the intrinsic time scale
around the critical temperature makes the time-scale matching condition
satisfied at two distinct temperatures, one below and the other above the
critical temperature, resulting in the double resonance peaks. The 
double SR peaks have also been observed in the classical Heisenberg spin
system in a planar thin film geometry~\cite{jang}, and infinite-range
$q$-state clock model~\cite{skbaek}.

\begin{figure}
\includegraphics[width=0.48\textwidth]{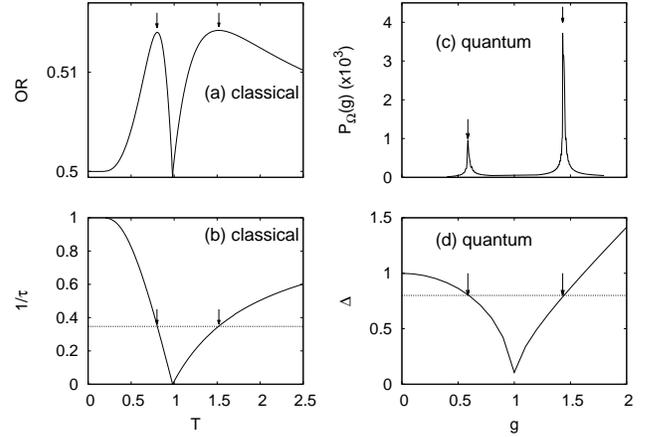}
\caption{Infinite-range (a)-(b) classical and (c)-(d) quantum Ising models.
Double SR peaks are observed in (a) 
the occupancy ratio (OR, the fraction of spins in the direction of the
external periodic driving), and  in (c) the power spectrum at the driving frequency 
$\Omega$, $P_{\Omega}(g) \equiv |\int dt e^{i\omega t} m(t; g)|_{\omega = \Omega}^2$,
where $m(t; g)$ is the magnetization.
Fluctuation strength is controlled by (a) the temperature $T$
and (c) the transverse field strength $g$, respectively. Arrows denote the 
positions of maxima in (a) and (c), which are in good agreements with 
the time scale matching conditions shown in (b) and (d).
In (b) $\tau$ is the intrinsic relaxation time scale, 
and in (d) $\Delta$ is the energy gap (and thus the
inverse time scale).  
The dotted horizontal lines in (b) and (d) denote the external 
frequency scales of time-periodic driving. $\Delta$ in (d) is obtained from
the direct diagonalization of the quantum Hamiltonian for the system size $N=1000$
and $P_{\Omega}(g)$ in (c)  through the semiclassical
approximation on the Heisenberg equation of motion (see text).
See~\cite{Kim} for (a) and (b).
}
\label{fig:cross}
\end{figure}

The SR phenomenon in quantum systems, named as 
quantum stochastic resonance (QSR), has been studied with focus on the interplay between 
quantum and classical fluctuation at finite temperatures~\cite{Gammaitoni,QSR}. 
The QSR at zero temperature has also been studied for the one-dimensional 
quantum spin system  with a spatially modulated external field, and the 
length-scale matching similar to the time-scale matching in conventional
SR has been discussed~\cite{Sen}.
In the present work, we study the QSR at zero temperature 
in the Ising spin system with the quantum phase transition~\cite{Sachdev}. 
We summarize our main findings in Fig.~\ref{fig:cross}, which displays 
the double SR peaks and the time-scale matching conditions in infinite-range 
classical~\cite{Kim} and quantum (this work) Ising 
systems in the presence of a weak external driving with the frequency $\Omega$. 
We conclude that the time-scale matching condition allows us to understand the 
classical and the quantum double SR peaks on the same ground.

In this paper, we numerically study the resonance behavior 
of the infinite-range quantum Ising model. 
Integrations of the time-dependent Schr\"odinger equation
and the semiclassical equation of motion unanimously yield
the existence of the double SR peaks, which are clearly
explained from the matching condition between
the energy gap, which is intrinsic, and the frequency of
the external time periodic driving.

\section{Results}
Let us begin with the globally coupled $N$ spins described by the Hamiltonian 
$H = -1/(2NS)\sum_{j\neq k}S^{z}_{j}S^{z}_{k}-g\sum_{j} S^{x}_{j}$,
where $S^{\alpha}_{j}$ is the spin-1/2 operator in the $\alpha$ direction ($\alpha=x, y, z$)
at the $j$th site ($S \equiv 1/2$ and $\hbar \equiv 1$ henceforth), and the transverse field $g$ 
in the $x$ direction 
induces quantum fluctuation due to $[S_{j}^{z},S_{k}^{x}]=i \delta_{jk} S_{j}^{y} \neq 0$.
By using the total spin operator $J_{\alpha}\equiv\sum_{j}S^{\alpha}_{j}$ 
with $J=N/2$, the Hamiltonian can be cast into the form~\cite{Botet} 
\begin{equation}
\label{eq:H}
	H = -\frac{1}{2J} {J_{z}}^{2}-g J_{x},
\end{equation}
which allows us to handle much bigger $N$ since the number of base kets 
becomes only $N+1$ (we use $J_z$ eigenkets as base kets).
The globally coupled quantum Ising model Eq.~(\ref{eq:H}) is very well-known
to exhibit the quantum phase transition of the 
mean-field nature and its finite-size scaling has also been
extensively studied~\cite{Botet} (see~\cite{Um} for the finite-size
scaling of the quantum phase transition in the one-dimensional Ising chain
system).

\begin{figure}
\includegraphics[width=0.48\textwidth]{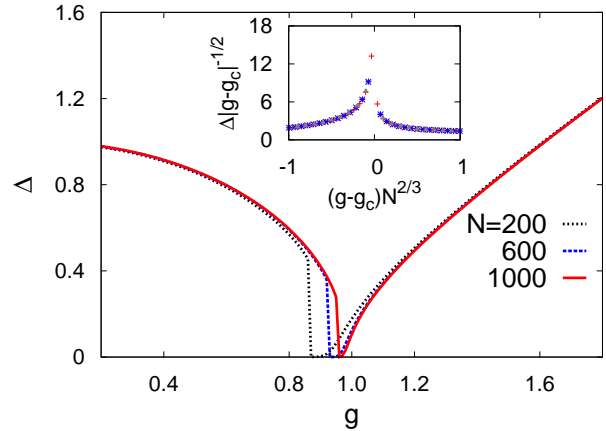}
\caption{(Color online) The energy gap $\Delta$ between the ground and the
first-excited states versus the transverse field strength $g$ for the globally-coupled
quantum Ising model without external driving in $z$-direction. 
$\Delta$ vanishes as the quantum critical point $g_{c}=1$ is approached. 
Inset: Finite-size scaling of the form 
$\Delta=(g-g_{c})^{z\nu}f[(g-g_{c})N^{1/d_c\nu}]$ with  
$z=1$, $\nu = 1/2$, and $d_c = 3$. 
}
\label{fig:gap}
\end{figure}

We numerically obtain the energy gap $\Delta$ between the ground and the first-excited states
of the Hamiltonian Eq.~(\ref{eq:H}),
which exhibits the quantum phase transition at $g=g_c = 1$ of the mean-field universality class 
as displayed in Fig.~\ref{fig:gap} for the system sizes $N=200, 600,$ and $1000$. 
The inset of Fig.~\ref{fig:gap} shows the finite-size scaling of $\Delta$ 
with the well-known exponents: dynamic critical exponent $z=1$, the correlation length exponent $\nu = 1/2$, 
and the upper critical dimension $d_c = 3$~\cite{Botet}. The vanishing energy gap (and thus
the divergence of the intrinsic time scale) at the quantum
critical point is particularly important in the present study: The nonmonotonicity of $\Delta$
as a function of the fluctuation strength $g$ provides the origin of the double quantum resonance
peaks (see Fig.~\ref{fig:cross}).

In parallel to studies of the classical SR behaviors~\cite{Gammaitoni,Kim}, we next apply 
the weak time-periodic external magnetic field $h(t)=h_{0}\cos\Omega t$ along the $z$-direction
with $h_0 = 10^{-3}$ and $\Omega = 0.8$,
to get the time-dependent Schr\"odinger equation 
\begin{equation}
\label{eq:Sch}
i\frac{d\vert \Psi(t) \rangle}{dt} = \left[ -\frac{1}{N} J_{z}^2-gJ_{x}-
h(t)J_{z}\right] \vert \Psi(t)\rangle,
\end{equation}
where the quantum ket $\vert \Psi(t)\rangle =\sum_{M=-J}^{J}A_M(t)
\vert M \rangle$ with $J_z|M\rangle = M|M\rangle$ and the complex coefficient $A_M(t)$.
The time evolution of the system is numerically traced through the use of
the fifth-order Runge-Kutta method combined with the Richardson extrapolation and 
Bulirsch-Stoer method~\cite{Press}. We check that the use of the sufficiently
small time step $\delta t = 10^{-4}$ keeps the normalization condition
$\sum |A_M(t)|^2 = 1$ unchanged within numerical accuracy.
We first get the ground state in the presence of 
the extremely small external field in the positive $z$-direction to break the
up-down spin symmetry, and use it as the initial condition for Eq.~(\ref{eq:Sch}).

\begin{figure}
\includegraphics[width=0.48\textwidth]{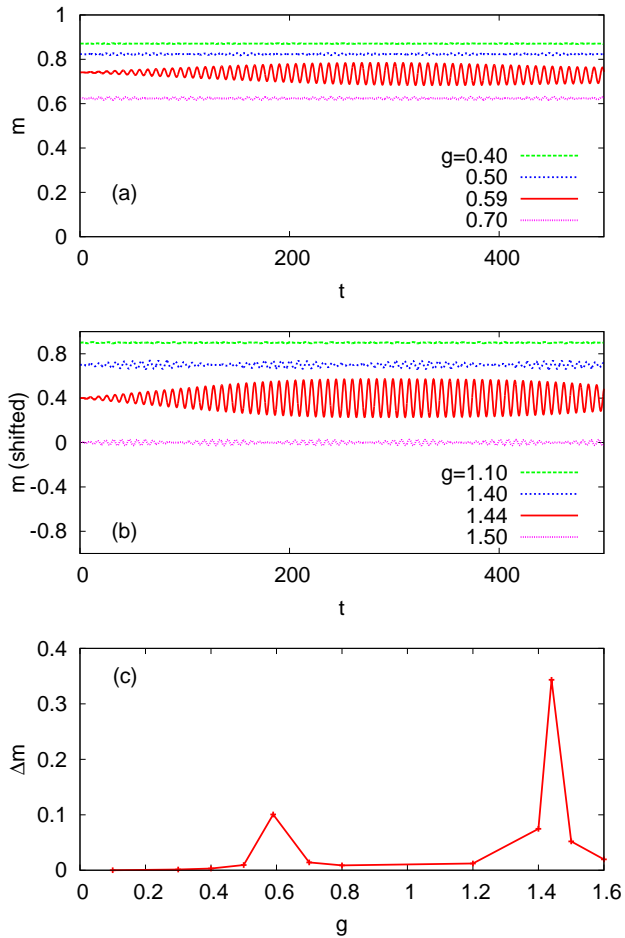}
\caption{
(Color online) The order parameter 
$m(t)$=$(1/N)\langle \Psi(t)\vert J_{z} \vert \Psi(t) \rangle$ in time $t$
at (a) $g < g_c = 1$ and (b) $g > g_c$ for $N$=$600$ in the presence of an external 
oscillating magnetic field $h_{0}\cos\Omega t $ with $\Omega$=$0.8$ 
in the $z$ direction.  At two values of $g$, one at $g_1\simeq 0.59$ and the other at $g_2\simeq 1.44$, 
$m(t)$ exhibits the larger oscillation amplitude reflecting the stronger coherence with the external
driving. For better distinction, $m(t)$ is vertically shifted by 0.9, 0.7, 0.4, and 0 at $g=1.10$, 1.40, 1.44, and
1.50 in (b), respectively. 
(c) The oscillation amplitude $\Delta m$ versus $g$, exhibiting double SR peaks. 
}
\label{fig:mt}
\end{figure}

\begin{figure}
\includegraphics[width=0.46\textwidth]{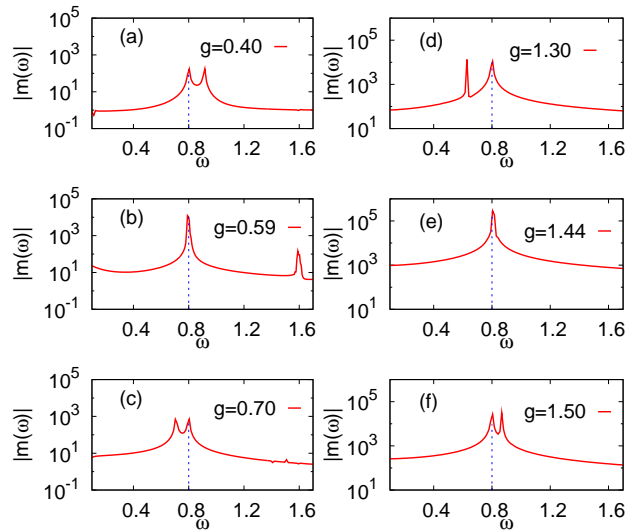}
\caption{
(Color online) The magnitude of the power spectral density 
 $|m(\omega)|$ versus frequency $\omega$ at various values of
the transverse field strength $g$.
The dashed vertical lines denote the frequency of the external field
($\Omega = 0.8$).  At $g$ = 0.59 and 1.44, the two peaks in the
spectral density merge to become one. 
The peak in (b) at $\omega$=1.6 is a harmonics, and does not reflect
the energy gap like peaks in other panels.
}
\label{fig:fourier}
\end{figure}

As the most important quantity to detect SR behavior, 
the average magnetization in the $z$ direction 
$m(t) \equiv (1/J) \langle \Psi (t) \vert J_{z} \vert \Psi (t) \rangle$ 
is measured as a function of time.
We do not observe significant difference for other system sizes, and 
we display our results $m(t)$ for $N=600$ in Fig.~\ref{fig:mt} at (a) below  and (b) above the 
quantum critical point $g_c = 1$. When $g > g_c$, $m(t)$ oscillates around $m = 0$, and 
we shift vertically each $m(t)$ in Fig.~\ref{fig:mt}(b) for better comparison. 
It is obvious from Fig.~\ref{fig:mt} that 
the resonance behavior of $m(t)$ is seen in the form of the larger oscillation amplitude 
at {\it two} distinct strengths of quantum fluctuation, i.e., one below $g_c$
and the other above $g_c$. 
In  Fig.~\ref{fig:mt}(c), we display the oscillation amplitude $\Delta m 
\equiv \max_t m(t) - \min_t m(t)$, which clearly shows double resonance peaks. 
We denote the first and the second resonance points
as $g_1 \approx 0.59$ and $g_2 \approx 1.44$, where the oscillation amplitudes
become maxima. 
As another indicator of the SR behavior, we carry out the Fourier transformation
of $m(t)$ to obtain $m(\omega)$ at frequency $\omega$.
Figure~\ref{fig:fourier} displays the magnitude of the spectral components 
$|m(\omega)|$ versus $\omega$ at various values of $g$. 
In general, there exist two peaks in $|m(\omega)|$, 
one at $\omega_1 = \Omega = 0.8$ [indicated by the dotted vertical lines in 
Fig.~\ref{fig:fourier}], and the other at the position $\omega_2$
that depends on $g$. We observe that the
latter peak at $\omega_2$ simply originates from the 
energy gap (see Fig.~\ref{fig:gap}), i.e., 
$\omega_2 = \Delta$ (note that $\hbar = 1$ in this work). 
As $g$ is increased toward $g_c$ from below,
$\Delta$ decreases (see Fig.~\ref{fig:gap}), which in turn
yields decreasing $\omega_2$ as shown in Fig.~\ref{fig:fourier}(a)-(c).
As $g$ passes through $g_c$ from below, $\omega_2$ bounces back and
moves to a larger value, reflecting the increase of $\Delta$ for $g > g_c$
in Fig.~\ref{fig:gap}. Another important observation one can make
from Fig.~\ref{fig:fourier} is that when the two peaks at $\omega_1$ and
$\omega_2$ merge into a single one at $\omega = \omega_1 = \Omega$, 
the power spectrum ($P_\omega = |m(\omega)|^2$) at $\Omega$ suddenly
increases much. From this, it is clear that the merging of the two
peaks must occur at two distinct values of $g$, which are in good
agreement with $g_1 \approx 0.59$ and $g_2 \approx 1.44$ in Fig.~\ref{fig:mt}.


\begin{figure}
\includegraphics[width=0.48\textwidth]{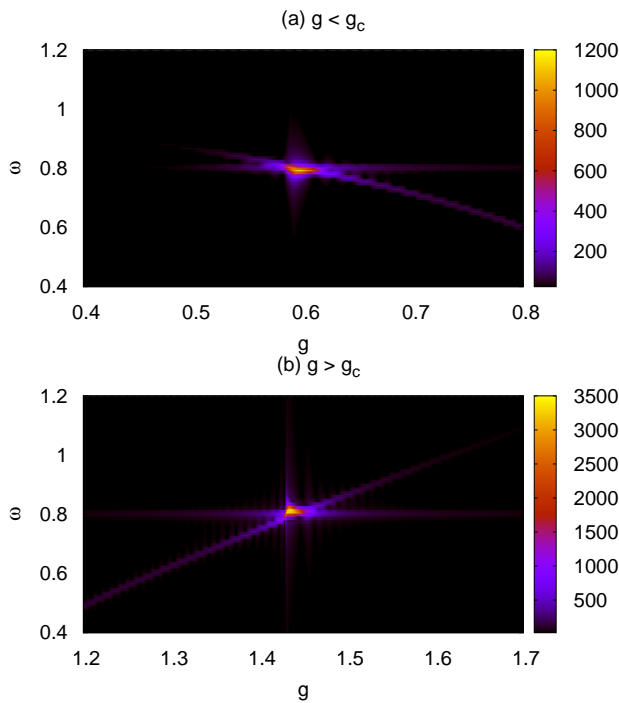}
\caption{(Color online) $|m(\omega)|$ versus $g$ and $\omega$
in the two regions (a) $g<g_{c}=1$ and (b) $g>g_{c}=1$ obtained via semiclassical 
approximation applied for the Heisenberg equations of motion. The right panels 
of (a) and (b) show the color scheme for $|m(\omega)|$.
Along $\omega=0.8$ axis, as $g$ is increased, 
$|m(\omega)|$  exhibits the biggest value 
both at $g\simeq0.59$ and $g\simeq1.44$. 
}
\label{fig:spectrum}
\end{figure}

We next adopt the Heisenberg picture in which the spin operator satisfies
the equation of motion $\dot{J_{\alpha}}=-i[H,J_{\alpha}]$.
By using the commutation relation $[J_{\alpha},J_{\beta}]$= 
$i\epsilon_{\alpha\beta\gamma}$$J_{\gamma}$, we get 
$\dot{J}_{x}=(1/2J)(J_{z}J_{y}+J_{y}J_{z})+h(t)J_{y}$, 
$\dot{{J}_{y}}=-(1/2J)(J_{z}J_{x}+J_{x}J_{z})+gJ_{z}-h(t)J_{x}$, 
and $\dot{{J}_{z}}=-gJ_{y}$. 
We then make the semiclassical approximation
and treat the spin operator $J_{\alpha}$ as the $\alpha$-th component 
of the classical spin vector 
$\vec{J}\equiv J(\sin\theta \cos\phi, \sin\theta \sin\phi, \cos \theta)$, 
which results in~\cite{Botet,Das} 
\begin{eqnarray}
\label{eq:semi}
\dot{\theta}&=&g\sin\phi, \nonumber \\
\dot{\phi}&=&g\cot\theta\cos\phi-\cos\theta - h(t).
\end{eqnarray}
We take initial values of $\theta$ and $\phi$ from 
the ground state of the system calculated by semiclassical 
approximation of Eq.~(\ref{eq:H}), and  perform numerical
integrations of Eq.~(\ref{eq:semi}) in time at given values of $g$.
In this semiclassical approximation, 
the order parameter is simply computed from $m(t) = J_{z}/J=\cos\theta(t)$,
which is then used for the Fourier analysis. 
We again find the SR behaviors at two distinct values of $g$:
$g_1\simeq0.59$ and $g_2\simeq1.44$ as displayed in Fig.~\ref{fig:spectrum},
which are in perfect agreement with the findings in Fig.~\ref{fig:fourier}.

\section{Summary}
In summary, the infinite-range quantum transverse-field Ising model at
zero temperature has been numerically
investigated in the presence of the weak longitudinal time-periodic magnetic
field at the frequency $\Omega$.
The resonance behavior at two distinct values of the transverse field $g$
has been clearly observed via (i) the large amplitude
oscillation of the magnetization in time and (ii) the large peak at $\Omega$ in spectral
analysis. The origin of the double SR peaks in the system has been identified from
the vanishing of the energy gap around the quantum critical point.
When the energy gap matches the frequency of the external field, the strong resonance 
peaks occur at two different values of $g$, exhibiting the double resonance behavior.
We have also confirmed the double resonance in the thermodynamic limit through the use of
the semiclassical approximation made for the Heisenberg equation of motion.
We propose that the time-scale matching condition should play an important role in understanding
the double SR behavior in a broad range of systems with continuous phase transitions,
both classical and quantum.

\section*{Acknowledgements}
This work was supported by the National Research Foundation of Korea (NRF) 
grant funded by the Korea government (MEST) via No. 2011-0015731.

\end{document}